\documentclass[%
reprint,
superscriptaddress,
amsmath,amssymb,  aps, prc, nofootinbib
]{revtex4-1}
\usepackage{mathtools}
\usepackage{graphicx}
\usepackage{dcolumn}
\usepackage{bm}
\usepackage{subfigure} 
\usepackage{xcolor}
\usepackage{ulem}

\begin{document}

\title{Fate of critical fluctuations in an interacting hadronic medium using maximum entropy distributions}

\author{Jan Hammelmann}
\thanks{Email: hammelmann@fias.uni-frankfurt.de}
\affiliation{Frankfurt Institute for Advanced Studies, Ruth-Moufang-Strasse 1, 60438 Frankfurt am Main, Germany}
\affiliation{Institute for Theoretical Physics, Goethe University, Max-von-Laue-Strasse 1, 60438 Frankfurt am Main, Germany}

\author{Marcus Bluhm}
\affiliation{SUBATECH UMR 6457 (IMT Atlantique, Nantes Universit\'e, IN2P3/CNRS), 4 rue Alfred Kastler, 44307 Nantes, France}

\author{Marlene Nahrang}
\affiliation{SUBATECH UMR 6457 (IMT Atlantique, Nantes Universit\'e, IN2P3/CNRS), 4 rue Alfred Kastler, 44307 Nantes, France}

\author{Hannah Elfner}
\affiliation{GSI Helmholtzzentrum f\"ur Schwerionenforschung, Planckstr. 1, 64291 Darmstadt, Germany}
\affiliation{Institute for Theoretical Physics, Goethe University, Max-von-Laue-Strasse 1, 60438 Frankfurt am Main, Germany}
\affiliation{Frankfurt Institute for Advanced Studies, Ruth-Moufang-Strasse 1, 60438 Frankfurt am Main, Germany}
\affiliation{Helmholtz Research Academy Hesse for FAIR (HFHF), GSI Helmholtz Center, Campus Frankfurt, Max-von-Laue-Straße 12, 60438 Frankfurt am Main, Germany}

\date{\today}

\begin{abstract}
We study the evolution of critical fluctuations in an expanding system within a hadronic transport approach.
The initialization of the system with critical fluctuations is achieved by coupling the ideal hadron resonance gas cumulants to the ones from the 3d Ising model and generating the net and total particle number distribution from the principle of maximum entropy. These distributions are then evolved using realistic hadronic interactions.
We systematically investigate the evolution of the critical fluctuations initialized at various temperatures and chemical potentials along a freeze-out line. We find that resonance regeneration and isospin randomization processes have the strongest influence on the evolution of the fluctuations.
Additionally, the sets of particles coupled to the critical mode are modified to assess the strength of the propagation of correlations through interactions.
We find that in the scaling region of the critical point correlations are propagated through the whole collisional history and are still present after the kinetic freeze-out of the matter if the coupling strength is large enough.
\end{abstract}
\maketitle

\section{Introduction}

  Studying the phase diagram of quantum chromodynamics (QCD) is one of the motivations in the field of heavy-ion collisions (HIC). With heavy nuclei accelerated in wide beam energies ranges HIC can probe different regions in temperature and baryon chemical potential. These experiments were performed e.g. at the HADES experiment at GSI \cite{HADES:2020wpc}, within the beam energy scan (BES) program which has continued with new results within the BESII program \cite{Kumar:2013cqa, STAR:2021iop} at the CERN SPS \cite{NA49:2009diu} or at the LHC \cite{ALICE:2022xpf}. 
  At vanishing baryon chemical potentials, lattice QCD calculations show that the phase transition between hadronic and partonic matter is a cross-over \cite{Aoki:2006we}. From effective models of QCD it is also known that a 1st order phase transition exists at larger baryon chemical potentials \cite{Fukushima:2010bq}. As a consequence there has to exist a critical point (CP) between those regions and the precise location is not known. 
  One of the observables that are thought to be sensitive to a possible CP are cumulants of conserved charges \cite{Asakawa:2000wh, Asakawa:2009aj, Hatta:2003wn}. In particular, the higher order cumulants exhibit a strong scaling behavior in the critical region. The largest impact is expected in the fluctuations of the baryon charge~\cite{Hatta:2003wn}. Therefore, the fluctuations of the net proton number, as a proxy for the net baryon number, are of great interest in both experimental and theoretical studies.
  
  Besides the challenges to interpret the experimental measurements, there also exist problems when it comes to modeling how to compare the results from theoretical calculations to the measurements. The theoretical description of fluctuations within the dynamical multi-stage evolution of the HIC is complicated and multiple questions must be addressed. First, the question arises of how the critical fluctuations form near the CP. This problem is typically tackled within a fluid dynamical framework in which the critical mode is treated as a field~\cite{Nahrgang:2011mg,Nahrgang:2011mv,Nahrgang:2014fza,Nahrgang:2018afz,Rajagopal:2019xwg,Nahrgang:2020yxm,Du:2020bxp,Du:2021zqz,Pihan:2022xcl}. The second question is then how such critical fluctuations survive the subsequent evolution in the dilute regime of the HIC towards the chemical and kinetic freeze-out. Our work contributes to the second question. To this end, we assume that critical fluctuations have built up at the moment where microscopic transport validly describes the hadronic phase. This occurs typically in the late stage of the collision and is, for noncritical dynamics, described by the Cooper-Frye particlization. We assume equilibrium fluctuations of the critical mode according to the three dimensional (3d) Ising model and couple it to a hadron resonance gas background. We then generate distribution functions of particle species using the principle of maximum entropy and evolve these distributions in a hadronic transport model. 
  Besides the question on the relation between the fluctuations of the net baryon and the net proton number which are measured in the experiment we can then tackle the question of how formations and decays of resonances alter the proton number fluctuations. 
  An approach based on a stochastic argument of isospin randomization processes which uses a binomial unfolding procedure was made in \cite{Kitazawa:2011wh, Kitazawa:2012at} to relate the proton and baryon fluctuations. The effect of primordial resonance decays on the net proton fluctuations was investigated in \cite{Fu:2013gga, Nahrgang:2014fza, Bluhm:2016byc}. To the best of our knowledge, no work has been done computing explicitly the effects of resonance regeneration processes on critical fluctuations.

  Regarding the first problem, what is known is the cumulants that originate from the coupling of the HRG model to the critical mode however, the resulting distribution function of some particle species is not known. Instead, it is the cumulants up to some finite order of these distributions and reconstructing a probability distribution from a finite set of moments or cumulants is a classical ill-posed problem in mathematics.
  In the appendix of \cite{Athanasiou:2010kw} a probability distribution was obtained by assuming the shape of the probability distribution of the sigma field $P(\sigma)$. However, the obtained distribution only approximately generated the correct input cumulants.
  To circumvent this problem an extension of the Cooper-Frye formula to include corrections of the critical field to a particle distribution function was performed up to second order in \cite{Pradeep:2022mkf}. In \cite{Pradeep:2022eil} the same authors made the general argument that the principle of maximum entropy is suitable for freezing out critical fluctuations.  
  In this work we want to follow the idea of \cite{Pradeep:2022eil} and convert fluctuations from a thermodynamic model that includes critical equilibrium fluctuations to particle spectra using the principle of maximum entropy. By doing so we obtain probability distribution functions of a specific particle species that we can put into a model for the dilute stage of a heavy-ion collision to evolve those particle spectra. We can then address the second-mentioned problem and investigate the impact of the stochastic nature of the hadronic phase on the fluctuations.

  The rest of this work is organized as follows. First, the description of the model is given with the definition of the fluctuations of the baseline, the inclusion of critical fluctuations and the reconstruction of the probability distribution using the maximum entropy method in Sec.~\ref{Sec:Model}. Then, the initial state of the simulations in terms of multiplicities and momenta of the hadrons are described in Sec.~\ref{Sec:IC} and then the transport model is introduced that is used to evolve the particle spectra in Sec.~\ref{Sec:SMASH}. 
  Sec.~\ref{Sec:Thermodynamics} starts with a description of the scatterings that take place in the medium and its thermodynamic evolution. In Sec.~\ref{Sec:Timeevol} the time evolution of the fluctuations will be presented and in Sec.~\ref{Sec:Isospin} the influence of isospin fluctuations is discussed. In Sec.~\ref{Sec:FinalState} the final state net proton and net nucleon fluctuations are shown and in Sec.~\ref{Sec:RapidityDependence} the rapidity dependence of the final state fluctuations is presented. This work is summarized in Sec.~\ref{Sec:Conclusion}.
  
\section{Model}\label{Sec:Model}

In this section, we discuss the baseline model which is used for the description of the fluctuations. First the hadron resonance gas (HRG) model is explained. Then, we discuss the coupling to the critical field which we relate to the 3d Ising model. Finally, the methodology for reconstructing the distribution functions of the net and total particle numbers is explained.

\subsection{Baseline model}
\label{Sec:hrg_baseline}

    As a baseline model for the description of the fluctuations we use the hadron resonance gas (HRG) model in the Boltzmann approximation. This implies that the equilibrium distribution function for a given particle species $i$ reads
    \begin{equation}
        \label{Eq:f_0}
        f^0_{i,k}(T, \mu_i) = \frac{1}{(2\pi\hbar)^3} \,\mathrm{exp}\left( -(E_{i,k}-\mu_i) / T\right)\,.
    \end{equation}
    Here, $T$ is the temperature of the system, $E_{i,k} = \sqrt{k^2+m_i^2}$ is the energy and $\mu_i = \mu_B B_i + \mu_Q Q_i + \mu_S S_i$ is the chemical potential of a particle of species $i$. From $f^0_{i,k}$ one obtains the particle density by integrating Eq.~(\ref{Eq:f_0}) over the three-momenta $\vec{k}$ as
    \begin{equation}
    \label{Eq:hrg_density}
        n_i(T, \mu_i) = \frac{e^{\mu_i / T}}{2\pi^2\hbar^3} m^2 T\, K_2(m_i/T) \,,   
    \end{equation}
    where $K_2$ denotes the modified Bessel function of the second kind. In the grand canonical ensemble (GCE), where the particle number is not fixed, the fluctuations in the particle number originate from the exchange with the thermal heat bath. The corresponding cumulants can be calculated by taking derivatives of $n_i/T$ with respect to $(\mu_i/T)$ at constant $T$. Given a fixed volume $V$, the mean number of particles in that system is $\kappa_{1,i} = N_i = V n_i$ and the higher order cumulants follow as
    \begin{equation}
    \label{Eq:cumulants_definition}
        \kappa_{n, i} = V T^3 \left.\frac{\partial^{n-1} (n_i / T^3)}{\partial(\mu_i / T)^{n-1}}\right\vert_T \,.
    \end{equation}
    One can see from calculating $\kappa_{n, i}$ with the density expression in Eq.~(\ref{Eq:hrg_density}) that the $n$-th order cumulant is given by the same expression as $\kappa_{1,i}$ meaning that the underlying probability distribution has to be of Poissonian nature. Since we are interested in the net and total particle and anti-particle numbers $N^\mathrm{net} = N^p-N^{\bar p}$ and $N^\mathrm{tot} = N^p+N^{\bar p}$ the cumulants in the HRG model are simply
    \begin{align}
        \label{Eq:HRGNet}
        \kappa_{n}^\mathrm{net} &= \kappa_{n}^p + (-1)^n \kappa_{n}^{\bar p}\,, \\
        \label{Eq:HRGTot}
        \kappa_{n}^\mathrm{tot} &= \kappa_{n}^p + \kappa_{n}^{\bar p} \,.
    \end{align}
    The relation in Eq.~(\ref{Eq:cumulants_definition}) allows us to relate a statistical measure such as a moment or a cumulant obtained within a theoretical framework to an experimental measurement.

    Cumulants are a statistical measure to quantify a given probability distribution similar to the moments. For a given distribution, one can calculate the cumulants by using the central moments $\mu_n = \langle (\delta N)^n \rangle$ with $\delta N = N - \langle N \rangle$ where $\langle\cdot\rangle$ denotes the average. The first three cumulants are identical with the same order central moments while the fourth order cumulant reads $\kappa_4 = \mu_4 - 3\mu_2^2$. One can relate any higher order cumulant of order $n>3$ to the central moments however we will focus on the first four cumulants in the following.
    
    From Eq.~(\ref{Eq:cumulants_definition}) one can see that the individual cumulants depend on the volume of the system. Therefore, neglecting any fluctuations in $V$, one usually shows ratios of different $\kappa_n$ in order to cancel the volume dependence, namely
    \begin{align}
        \sigma / M &= \frac{\kappa_2}{\kappa_1} \,\\
        S\sigma &= \frac{\kappa_3}{\kappa_2} \,\\
        \kappa\sigma^2 &= \frac{\kappa_4}{\kappa_2} \,.
    \end{align}
    These ratios are known as the scaled variance, skewness and kurtosis, respectively.

\subsection{Critical mode fluctuations}
\label{Sec:CritModeFlucs}
  
    Since the HRG model is a model without any criticality we need to extend our baseline model in order to incorporate critical behavior. By using the universality class argument~\cite{Hohenberg:1977ym} that in the vicinity of the critical point the scaling behavior of the critical order parameter in QCD is the same as for the order parameter in the 3d Ising model, we can relate the fluctuations in the order parameter of the chiral phase transition with those in the order parameter of the 3d Ising model, the magnetization.

    The model we employ here to describe the critical fluctuations is the same one that has been used in~\cite{Bluhm:2016byc}. We therefore only briefly describe it here for the sake of completeness. The equation of state (EoS) for the 3d Ising model is taken from a parametric representation of the magnetization that reproduces faithfully the scaling behavior with the critical exponents~\cite{Guida:1996ep}:
    \begin{align}
        \label{Eq:Ising_M}
        M &= M_0 R^\beta \theta \,,\\
        \label{Eq:Ising_r}
        r &= R (1 - \theta^2) \,,\\
        \label{Eq:Ising_h}
        h &= H / H_0 = R^{\beta \delta} \tilde h(\theta) \,.
    \end{align}
    Here, $r = (T - T_c) / T_c$ and $h = H / H_0$ are the reduced temperature and the reduced external magnetic field, respectively. $R$ and $\theta$ are auxiliary variables, $M_0$ is a normalization constant and $\tilde h(\theta) = c\theta (1 + a\theta + b\theta^4)$. The universal scaling is built into the EoS with the critical exponents $\beta = 0.3250$ and $\delta = 4.8169$, and the coefficients in $\tilde h(\theta)$ read $a = -0.76145$, $b = 0.00773$ and $c = 1$. Similar to Eq.~(\ref{Eq:cumulants_definition}), the fluctuations of the order parameter field $\sigma$ can be calculated via derivatives with respect to the reduced magnetic field as
    \begin{equation}
      \langle (\delta\sigma)^n\rangle_c = \left( \frac{T}{VH_0}\right)^{n-1} \left.\frac{\partial^{n-1}M}{\partial h^{n-1}}\right|_r \,.
    \end{equation}
    The rather lengthy expressions for the cumulants up to the fourth order can be found in~\cite{Bluhm:2016byc}. The mapping between the 3d Ising model and QCD is then performed by relating the reduced Ising temperature and external magnetic field with the temperature and the baryon chemical potential in the QCD phase diagram. An additional rotation of $r$ by an angle $\alpha$ is introduced and $h$ is taken to be parallel to the temperature axes via
    \begin{align}
        \frac{T - T_c}{\Delta T_c} &= r\, \mathrm{sin}\,\alpha_1 \,,\\
        \frac{\mu_B - \mu_{B,c}}{\Delta\mu_{B,c}} &= -r\, \mathrm{cos}\,\alpha_1 - h \,.
      \end{align}
    The temperature of the QCD critical point $T_{c}$ and the angle $\alpha$ are obtained by assuming that the critical point lies on the phase transition line determined in~\cite{Borsanyi:2010bp, Cea:2015cya, Endrodi:2011gv}. The value we choose for the critical point is $\mu_{B,cp} = 0.39$~GeV which sets $T_{cp} \approx 0.148$~GeV. The scaling sizes are chosen to be $\Delta T_{c} = 0.02$~GeV and $\Delta \mu_{B,c} = 0.42$~GeV.
    
    \begin{figure}[ht]
      \centering
      \includegraphics[width=0.5\textwidth]{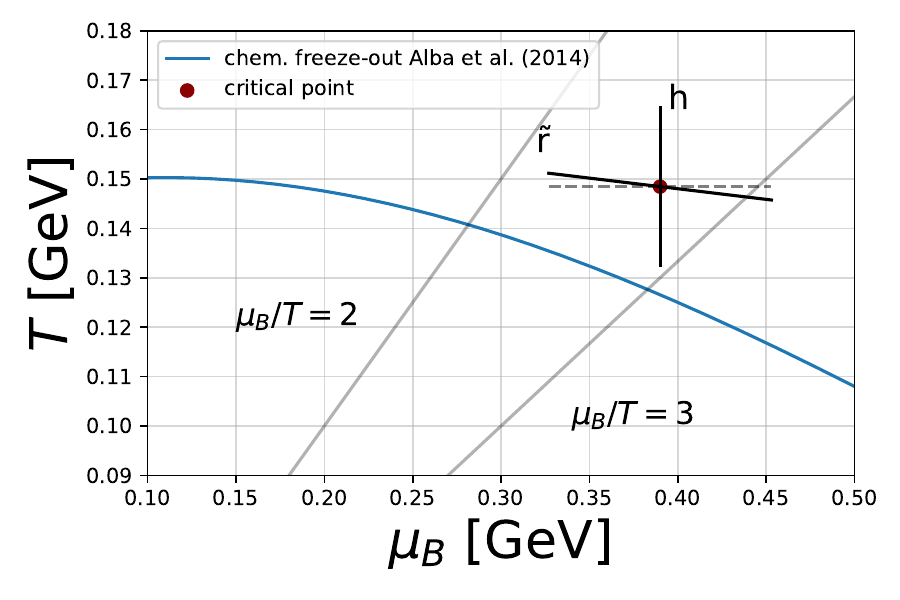}
      \caption{Sketch of the QCD phase diagram showing the location of the QCD critical point used in this work and a parametrization for the chemical freeze-out curve~\cite{Alba:2014eba}.}
      \label{Fig:qcd-pd}
    \end{figure}
    A sketch of the QCD phase diagram including a critical point and the positioning of the $r$ and $h$ axes is shown in Fig.~\ref{Fig:qcd-pd}. QCD effective theory calculations predict that the QCD critical point is located at rather large $\mu_B/T \gtrsim 4$ (see for example the functional renormalization group calculations in~\cite{Fu:2019hdw,Gao:2020qsj,Fu:2021oaw} or the Dyson Schwinger formalism computations in~\cite{Fischer:2018sdj,Isserstedt:2019pgx}). However, since we do not aim for a direct comparison with experimental measurements we employ the parameters that have been used in~\cite{Bluhm:2016byc}.
    
    In order to couple the distribution function in the HRG model to the critical mode one needs a suitable formalism. Here, we use the ansatz made for example in~\cite{Stephanov:1999zu,Stephanov:2011pb} where the equilibrium distribution function is extended by an additional critical contribution $\delta f_i^{\mathrm{critical}}$ via 
    \begin{equation}
        f_{i, k} = f_{i, k}^0 + \delta f_i^\mathrm{critical} \,.
    \end{equation}
    The critical contribution originates from the coupling of the $\sigma$ field to the particles of species $i$ which dynamically influences their masses via $\delta m_i = g_i\delta \sigma$ and one can therefore write
    \begin{equation}
    \label{Eq:delta_f}
      \delta f^{\mathrm{critical}}_i = -\delta\sigma \frac{g_i}{T}\frac{f_{i,k}^0}{\gamma_{i,k}} \,.
    \end{equation}
    Here, $\gamma_{i,k} = E_{i,k} / m_i$ and $g_i$ denotes the strength of the coupling which will be used as a free parameter in this work.
    This procedure allows us to derive the cumulants of the individual particle and anti-particle numbers including contributions from critical fluctuations~\cite{Bluhm:2016byc}.

    As we will see below, in order to generate discrete samples of particles and anti-particles of a specific species it is necessary to know the expressions for the total and the net particle number cumulants. The critical contributions induce correlations between particles and antiparticles and therefore Eqs.~(\ref{Eq:HRGNet}) and~(\ref{Eq:HRGTot}) have to be extended in order to account for those. In compact form, we write the modification of the net and total particle number cumulants of a particle $p$ and its anti-particle $\bar p$ for $n>2$ (the mean number is not affected by the critical contributions) as
    \begin{align}
        \kappa_n^{\mathrm{net}} &= \kappa_n^p + (-1)^n \kappa_n^{\bar p} + (-1)^n \langle (V\delta\sigma)^n\rangle_c (I_p - I_{\bar p})^n \,,\\
        \kappa_n^{\mathrm{tot}} &= \kappa_n^p + \kappa_n^{\bar p} + (-1)^n \langle (V\delta\sigma)^n\rangle_c (I_p + I_{\bar p})^n \,,
    \end{align}
    where
    \begin{equation}
      I_i = \frac{g_i d_i}{T}\int \frac{d^3k}{(2\pi\hbar)^3} \frac{f_{i,k}^0}{\gamma_{i,k}} \, .
    \end{equation}

    Now, we are able to describe the effect of the critical point on the net and total particle number cumulants for a given particle species $i$. However, since the transport model that we use evolves the single-particle distribution functions, a methodology is needed to translate the cumulants into samples of (anti-)particles which can be evolved in the transport model.

\subsection{Maximum entropy method for freezing out critical fluctuations}
\label{Sec:MaxEntropyMethod}
    
    Since the probability distribution of a particle species coupled to the critical mode is unknown, we have to construct $P(N_i)$ from the known cumulants. However, the reconstruction of a probability distribution is a mathematically ill-posed problem as one needs all cumulants (or moments) in order to define the distribution. One way around this problem is to impose an additional criterion. Here, we use as the criterion that the information entropy of the distribution is maximized since it is reasonable to assume that distributions that are realized in nature have a maximum information entropy~\cite{Mead:1983qg}. It was also realized by Jaynes that there exists a fundamental mathematical equivalence between the principle of maximum information entropy and statistical mechanics~\cite{Jaynes:1957zza}.

    The Shannon information entropy (which is also equivalent to the entropy of the canonical ensemble up to a factor of $k_B$) of a discrete probability distribution $P(x)$ is given by
    \begin{equation}
      S = - \sum_{x\in\Omega} P(x)\, \mathrm{ln}\,P(x) \,,
    \end{equation}
    where $\Omega$ is the support of $P(x)$. By enforcing $S$ to be maximized under the condition that the moments of $P(x)$ are of a specific value, one can derive the expression of the maximum entropy (ME) distribution $P_\lambda(x)$ for the Lagrange-multiplier $\lambda$ as 
    \begin{equation}
    \label{Eq:MEDist}
      P_\lambda(x) = Z^{-1}_{\lambda}\mathrm{exp}\left\{ \sum_{k=1}^{n} \lambda_k x^k \right\} \,.
    \end{equation}
    Here, the partition function is defined as $Z_\lambda = \sum_{x\in\Omega}\mathrm{exp}\left\{ \sum_{k=1}^{n} \lambda_k x^k \right\}$ which is used in order to normalize $P_\lambda(x)$.
    Now, with the input of $n$ moments from the critical model one obtains the following $n$ equations for the moments of the maximum entropy distribution function $P_\lambda(x)$
    \begin{align}
        \mu_k = -\frac{\partial \mathrm{log} Z_{\lambda}}{\partial \lambda_k} = Z_{\lambda}^{-1}\tilde\mu_k \, ,
    \end{align}
    with $\tilde\mu_k = \sum_x x^k\mathrm{exp}\left\{ \sum_{l=1}^{n} \lambda_l x^l \right\}$.
    Together with the Jacobian $J_{n,m} = Z_\lambda^{-2}\tilde\mu_m\tilde\mu_n - Z_\lambda^{-1}\tilde\mu_{m+n}$ the $n_m$ equations are solved using a Newton method to obtain the Lagrange multipliers $\lambda_k$.

    The obtained distribution function is not the exact physical distribution function of the particle number. However, it is the one with the maximum amount of uncertainty. Given that currently only the first four cumulants from the 3d Ising model are available with parametrizations given in~\cite{Bluhm:2016byc}, the maximum entropy distribution is the least biased estimate of the critical particle number distribution that one can construct with the given information.
    
    It is worth mentioning that the particle number distribution from the HRG model follows a Poisson distribution which itself is a distribution of maximum entropy. Therefore, in the limit $g_i\rightarrow 0$ for the coupling of particles of species $i$ to the critical mode the baseline distribution is restored. On the other hand, it is not possible to go to infinite values of the coupling strength with the current method. At some point, we find numerical difficulties in obtaining reasonable values for the Lagrange multipliers in the sense that the probability distribution is physical on the support $\Omega$.

\section{Initial state}\label{Sec:IC}
\label{Sec:InitialState}

    This section describes the modeling of the initial state of the simulation in both momentum and coordinate space.
    We want to initialize the transport model with critical equilibrium fluctuations to study their evolution and we therefore need the initial phase space information of the particles which we call the initial state.
    In this work, we assume a simplified geometry at particlization. More realistic scenarios for HIC will be discussed in future studies. The first simplification is that the hadrons are sampled uniformly within a sphere of radius $R$. We take all available hadrons from SMASH (see {\tt particles.txt} \cite{SMASH_github}).
    The momentum space of the particles is sampled from the modified Boltzmann distribution
    \begin{equation}
        f_{i,k} = e^{- u\cdot k_i / T} \, .
    \end{equation}
    Here $\vec u(r) = \vec e_r u_0 r / R$ is a velocity field that has been introduced to reproduce experimentally measured momentum distributions \cite{Ling:2015yau, STAR:2017sal}. 
    The temperature and chemical potentials in the initial state are obtained from the freeze-out curve \cite{Alba:2014eba} which maps the beam energy $\sqrt s$ of a HIC to equilibrium values of temperature $T$ and chemical potentials $\mu_{Q, B, S}$. This estimate is based on the analysis of net-electric charge and net-proton number fluctuations of experimental measurements. The parametrization used in this work can be found in \cite{Bluhm:2016byc}.
    \begin{figure}[h]
      \centering
      \includegraphics[width=0.4\textwidth]{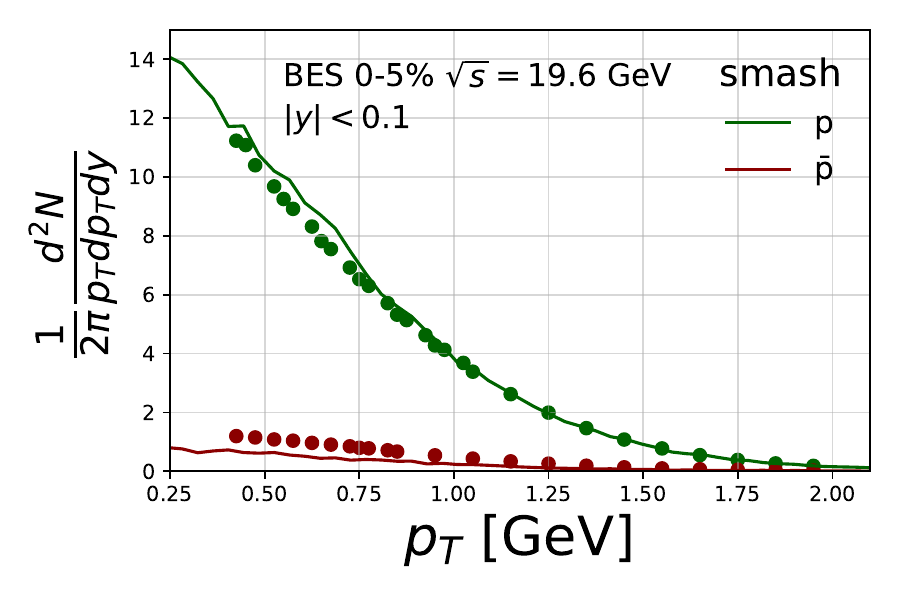}
      \caption{$p_T$ spectra of protons (green) and anti-protons (red) at $\sqrt{s} = 19.6$ GeV with $u_0 = 0.5$ and $R_{\mathrm{sphere}} = 9$ fm. Experimental data points are taken from \cite{STAR:2017sal}}
      \label{Fig:spectra}
    \end{figure}
    Fig. \ref{Fig:spectra} shows the $p_T$ spectra of protons and anti-protons measured from STAR at a beam energy of $\sqrt s = 19.6$ GeV in midrapidity \cite{STAR:2017sal}. It shows that with an expansion parameter $u_0 = 0.5$ the slope of the $p_T$ spectra can be well described. For simplicity and throughout this work a value of $R=9\,\mathrm{fm}$ and $u_0 = 0.5$ is used across the energy scan region $\sqrt s = 7 - 50$ GeV.

    In the case of coupling the particle distribution to the critical mode, we first generate samples of the net and total numbers $N^{\mathrm{Net / Tot}}$ from the ME distributions by sampling their cumulative distribution function. Since the transport model evolves the single-particle distribution function, one needs to generate samples of the particle and anti-particle numbers respectively. As not all pairs of net and total particle numbers are valid in the sense that they result in positive integer-valued numbers of particles, one still has to find valid pairs in the $\{N^{\mathrm{Net/Tot}}\}$ samples. This is achieved by simply searching for valid combinations within the two samples $\{N^{\mathrm{Net}}\}$ and $\{N^{\mathrm{Tot}}\}$ and removing them from the samples.
    \begin{figure}[h]
      \centering
      \includegraphics[width=0.45\textwidth]{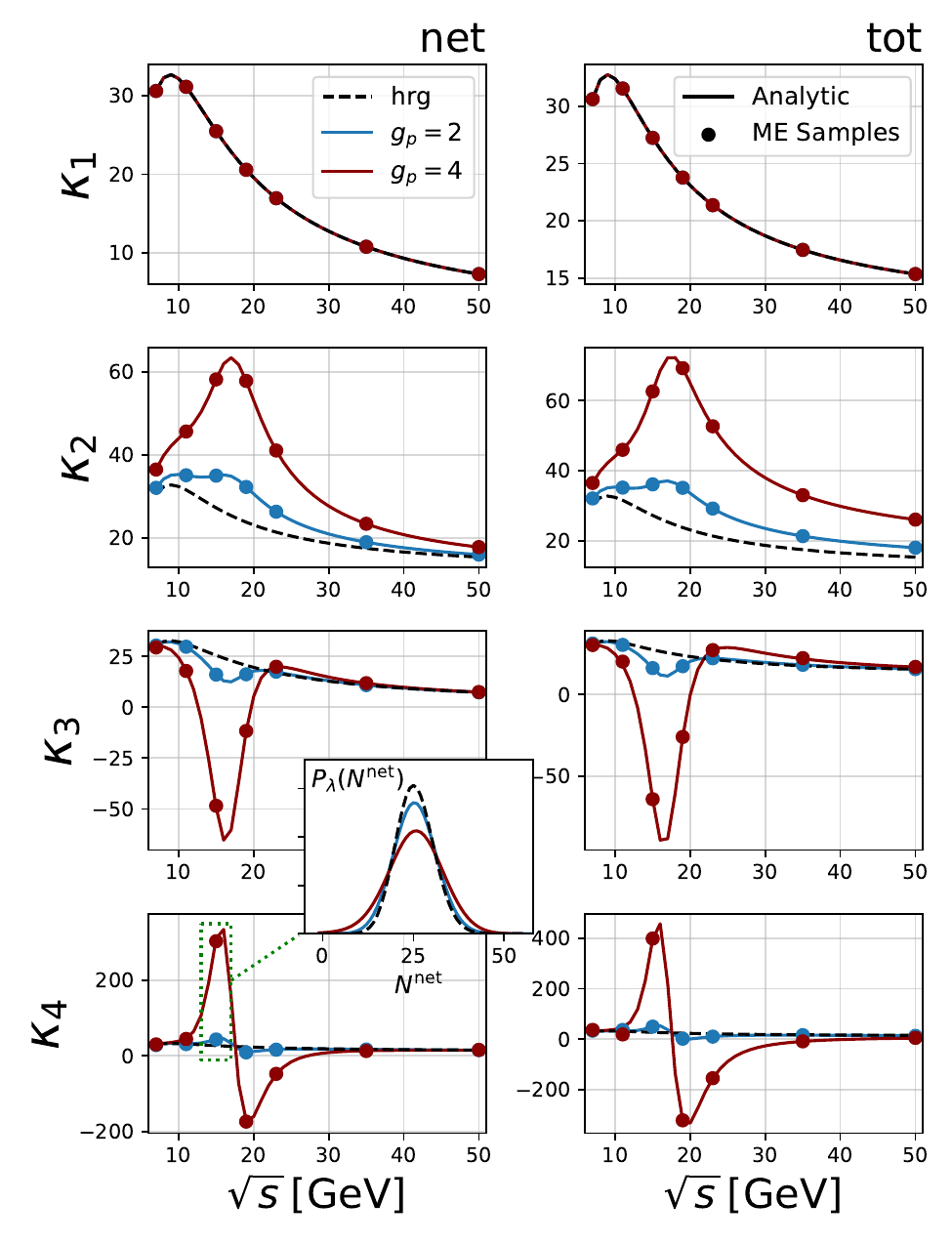}
      \caption{Net (left column) and total (right column) proton number cumulants as a function of $\sqrt s$ up to fourth order (from top to bottom). The analytic HRG baseline (grey dashed) as well as the calculation including the critical mode with a coupling of $g_p = 2, 4$ (full blue and red) are presented. In addition, the results from the generated particle samples are presented as red circles. In addition, the reconstructed net proton number ME distributions are shown for $\sqrt{s} = 15\,\mathrm{GeV}$ plus the skellam distribution in the small subplot.}
      \label{Fig:ProtonCumulantsSample}
    \end{figure}

    Fig. \ref{Fig:ProtonCumulantsSample} shows the net and total proton number cumulants up to fourth order along the freeze-out line. The results of the ME samples were computed from the proton and anti-proton samples for a coupling of $g_p = 2$ and $g_p = 4$ to the critical field.
    This result shows that our methodology successfully reproduces simultaneously the net and total proton cumulants up to fourth order at a given value of $\sqrt{s}$.

    As an example, the distribution function $P_\lambda(N^{\mathrm{net}})$ for $\sqrt{s} = 15\,\mathrm{GeV}$, see inlay of Fig.~\ref{Fig:ProtonCumulantsSample}, shows the modification of the critical fluctuation to the skellam distribution. For the coupling strength of $g_p = 2$ the distribution function only slightly differs from the baseline curve. With increasing coupling strength, the tails of the distribution start to grow and the width increases.
    We have chosen two values of the critical coupling strength ($g_p = 2,4$) in order to vary the magnitude of the equilibrium critical contributions. Further, we distinguish between two cases. In the first case only (anti-)protons and (anti-)neutron couple to the critical field. For the second case we include a larger set of hadronic species which are coupled to the critical mode. For more details, see App.~\ref{App:2}. We use the phenomenological approach to determine the respective critical coupling strength \cite{Bluhm:2016byc}
    \begin{equation}
      g_R = \frac{g_c}{N_q}\frac{m}{m_R}(N_q - |S_R|) \, .
    \end{equation}
    Here $N_q$ is the number of valence quarks of the hadron, $g_c$ and $m$ the coupling and mass of the respective stable hadron and $|S_R|$ the absolute strangeness number. As our methodology of determining the critical probability distribution faces numerical issues when going to large values of the coupling strength $g$, we are restricted to $g_p = 2$ for other hadronic species than nucleons.
    To summarize, we are going to study three different coupling scenarios which are:
    \begin{itemize}
        \item Couple only nucleons to the critical field using $g_c = 2,4$
        \item Couple particle species listed in App.~\ref{App:2} with a baryonic and mesonic coupling of $g_{c} = 2$ (denoted as $g_c = 2+$).
    \end{itemize}
    With these three cases, we are able to study the influence of the coupling strength as well as the case in which more hadron species are coupled to the critical field.
    Finally, we note that we do not include any modifications in momentum space. It is also expected that the fluctuations of momenta are modified due to the presence of the critical field.

    In the next step, we want to take these distributions for each $\sqrt s$ and evolve them microscopically in the hadronic transport approach to see how they get affected. 

\section{SMASH}\label{Sec:SMASH}
    The evolution of the hadronic medium is modeled with the hadronic transport approach SMASH \cite{Weil:2016zrk,dmytro_oliinychenko_2020_4336358,SMASH_github}. 
    The model has been used successfully to simulate HIC at various different collisional energies including a hybrid model \cite{Mohs:2019iee, Staudenmaier:2020xqr, Schafer:2021csj} but also to extract bulk properties of the hadronic medium like the shear or bulk viscosity or diffusion coefficients \cite{Rose:2017bjz, Rose:2020lfc, Hammelmann:2023fqw}. 

    Hadronic transport approaches are built to evolve the single-particle distribution function and the interactions are performed based on the input of experimentally measured cross-sections. Typically, in such models a criterion is needed in order to decide if a collision happens. In this work, we employ a covariant form of the geometric collision criterion \cite{Hirano:2012yy} which uses the geometric interpretation of the cross-section
    \begin{equation}
        d_\perp < \sqrt{\sigma_{\mathrm{tot} / \pi}} \, .
    \end{equation}
    Here, $d_\perp$ is the transverse distance between two hadrons and $\sigma_{\mathrm{tot}}$ the total cross-section of the reaction.

    Transport codes yield an effective solution of the relativistic-Boltzmann equation \cite{Tindall:2016try}. However studying fluctuations within the Boltzmann equation itself is in principle not possible since the information about the distribution function $f_{i,k}$ is only of probabilistic nature. In a transport code like the one we use however the single particle particle distribution function is evolved on a Monte Carlo basis and the solution to the Boltzmann equation is given by event averages. This enables one to effectively study fluctuations since each particle's phase space information is accessible at each point in phase space.

    The geometric collision criterion limits the number of reaction partners to two so the implemented interactions are resonance formation and decay processes $2\leftrightarrow 1$ and in-/elastic interactions $2\leftrightarrow 2$. As the resonance formation and decay interactions play an important role, we want to explain their treatment more thoroughly.
    The decay widths are implemented as mass-dependent $\Gamma(m)$ following the idea of \cite{Manley:1992yb} and the spectral function of the resonance uses the relativistic Breit-Wigner distribution with mass-dependent widths. However, during the evolution the decay widths feel an effective broadening \cite{Hirayama:2022rur}. The cross-section of forming such a resonance is then given by
    \begin{equation}
        \sigma_{ab\rightarrow R}(s) = \frac{2 J_R + 1}{(2 J_a + 1)(2 J_b + 1)}S_{ab}\frac{2\pi^2}{\vec p_i^2} \Gamma_{ab\rightarrow R}(s) \mathcal{A}_R(s)
    \end{equation}
    Here $J$ is the spin of the particle, $\vec p_i^2$ the center of mass momentum of the reaction, $S_{ab}$ a symmetry factor (2 if $a = b$ and 1 if $a\neq b$), $\Gamma(s)$ is the partial width of the resonance and $\mathcal{A}$ its spectral function.
    The total cross-section of the reaction between a particle $a$ and particle $b$ is then
    \begin{equation}
        \sigma_{\mathrm{tot}}^{ab}(s) = \sum_R \sigma_{ab \rightarrow R}(s) + \sum_{x, y} \sigma_{ab \rightarrow xy}(s) + \sigma_{ab \rightarrow \mathrm{string}}(s) \, ,
    \end{equation}
    with the cross section of a (in)elastic $2\rightarrow 2$ reaction $\sigma_{ab \rightarrow xy}$ and the formation of a string $\sigma_{ab \rightarrow \mathrm{string}}$ \cite{Mohs:2019iee}. We note that the formation of strings via the so-called yo-yo model plays no large role at the considered energies.
    This approach is a bottom-up approach where the sum runs over all possible particles in the particle list to fit the total cross-section with experimental measurements.

    When a resonance $R$ is formed it propagates and, based on a given time step $\Delta t$ of the simulation it will decay with the following probability
    \begin{equation}
        P(\mathrm{decay\, at\, \Delta t}) = \Gamma(m)\Delta t \, .
    \end{equation}
    If multiple possible decay channels exist, one of them is chosen from the probability $p_i = \Gamma_i(m) / \Gamma(m)$. The momenta of the outgoing particles are sampled isotropic.

    Within this work, we don't employ any potentials so the evolution is only performed in the so-called cascade mode. This leaves possible work for the future to incorporate dynamics of a critical point into the transport code via potentials.

\section{Results}
  \subsection{Thermodynamic evolution and collision chemistry of the medium}\label{Sec:Thermodynamics}
    We first want to study the thermodynamic properties of the expanding sphere by determine the temperature and baryon chemical potential after the evolution of the medium.
    \begin{figure}[h]
      \centering
      \includegraphics[width=0.5\textwidth]{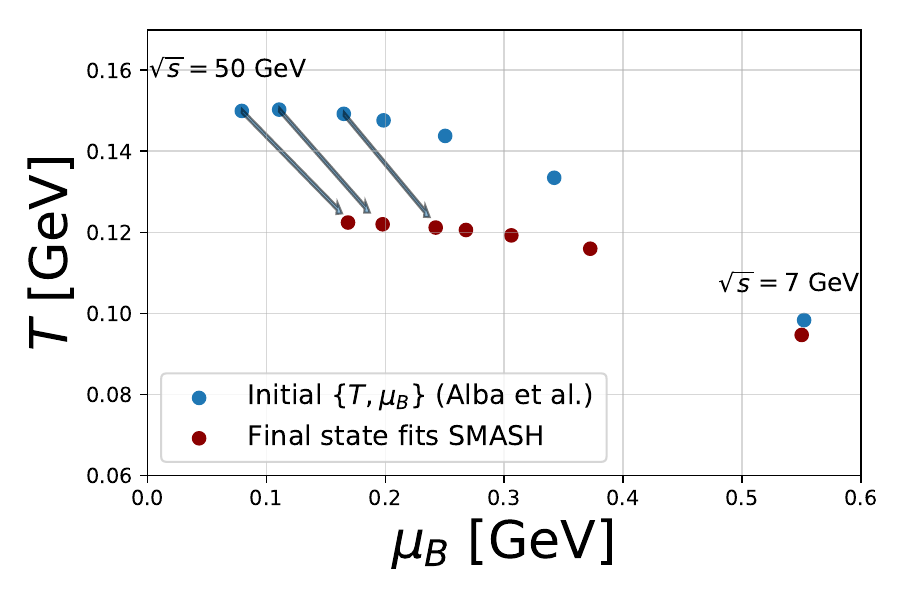}
      \caption{Evolution of the temperature and baryon chemical potential of the expanding medium. The initial state values are shown as blue and the final state values as red points.}
      \label{Fig:final-state_T-mu}
    \end{figure}
    The thermodynamic evolution of the medium in terms of temperature and baryon chemical potential is shown in Fig.~\ref{Fig:final-state_T-mu}. For each point along the freeze-out curve a thermal model fit is performed on the final state of the evolution to obtain the equilibration values of the temperature, the chemical potentials as well as the volume. The details are explained in App.~\ref{App:1}.
    The final state is obtained by evolving the system until no interactions are occurring anymore and all resonances that are still present in the simulation are decayed into the ground states. We find that the kinetic freeze-out is reached at approximately $t = 100\,\mathrm{fm}$, note that this is not a realistic average time for freeze-out in the full collision, but the final time after the last interaction in our simplified spherical geometry occurs.

    As the hadron gas including its resonances is initialized according to the temperature and chemical potentials from the freeze-out curve the final thermodynamic values are expected to change as the chemical composition of hadron gas changes.
    We find that the temperature of the system decreases and the baryon chemical potential increases for the systems initialized at large $\sqrt{s}$ energies due to the decay of heavy resonances.
  
    In the next step, we want to investigate the collision chemistry of the expanding hadronic system. By doing so we can directly observe which types of interaction are the most important ones during the evolution.
    \begin{figure}[h!]
      \centering
      \includegraphics[width=0.5\textwidth]{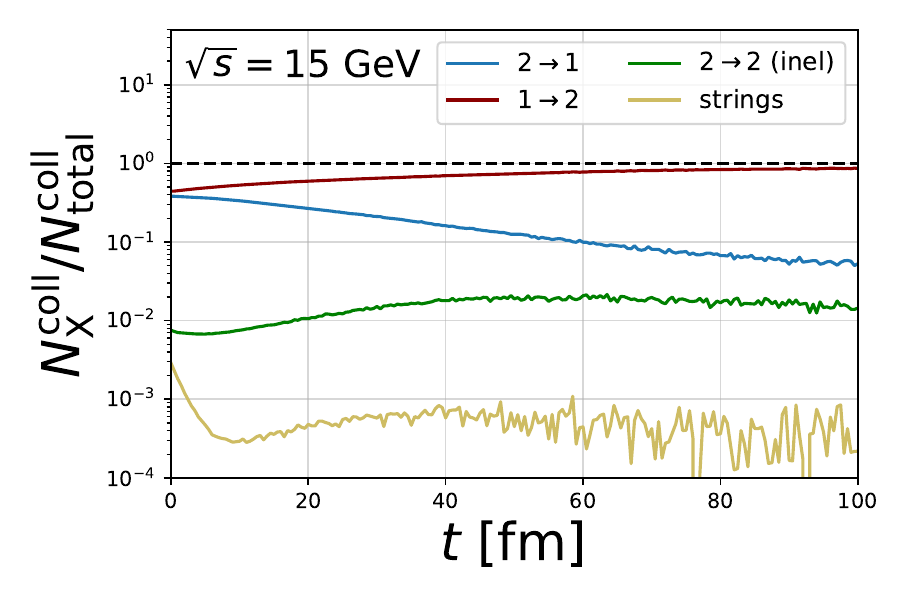}
      \caption{Number of resonance decays (red), resonance formations (blue), inelastic $2\rightarrow 2$ (green) and string formation (yellow) normalized to the total amount of collisions as a function of time.}
      \label{Fig:collisions}
    \end{figure}
    Fig.~\ref{Fig:collisions} shows the number of specific collision channels as a function of time. In this example we have picked an energy on the freeze-out curve of $\sqrt{s} = 15\,\mathrm{GeV}$. It has been checked that similar results are obtained with other energies.
    From Fig.~\ref{Fig:collisions} one can see that during the whole evolution, the resonance decays, followed by resonance formations are the most dominant types of interaction. Inelastic $2\rightarrow 2$ scatterings are on the order of $\sim 1\%$ and finally the string formation processes play only a subdominant role during the expansion.
    
    In \cite{Fu:2013gga}, it has been derived how resonance decays affect particle number cumulants. The decay chain of unstable resonances into stable particles yields a source of fluctuations effectively increasing the cumulants. In the system studied here, an additional source of fluctuations exists as resonances can be newly formed resulting in a non-trivial interplay between formation and decay.
    In addition to the generation of resonances, they can also be created in inelastic $2\rightarrow 2$ collisions such as $NN\rightarrow NN^\star$ even though these reactions are of sub leading order they do have an effect on the net proton number since they, first, reduce the net proton number at the time of the collision and second, randomize the isospin in the time of the decay of the resonance which has additionally modifies the net proton cumulants.

  \subsection{Time evolution}\label{Sec:Timeevol}
    We now want to discuss the evolution of the individual cumulants as a function of time. By doing so we are able to distinguish the impact of different phases of the evolution on the fluctuations.
    \begin{figure}[h!]
      \centering
      \includegraphics[width=0.5\textwidth]{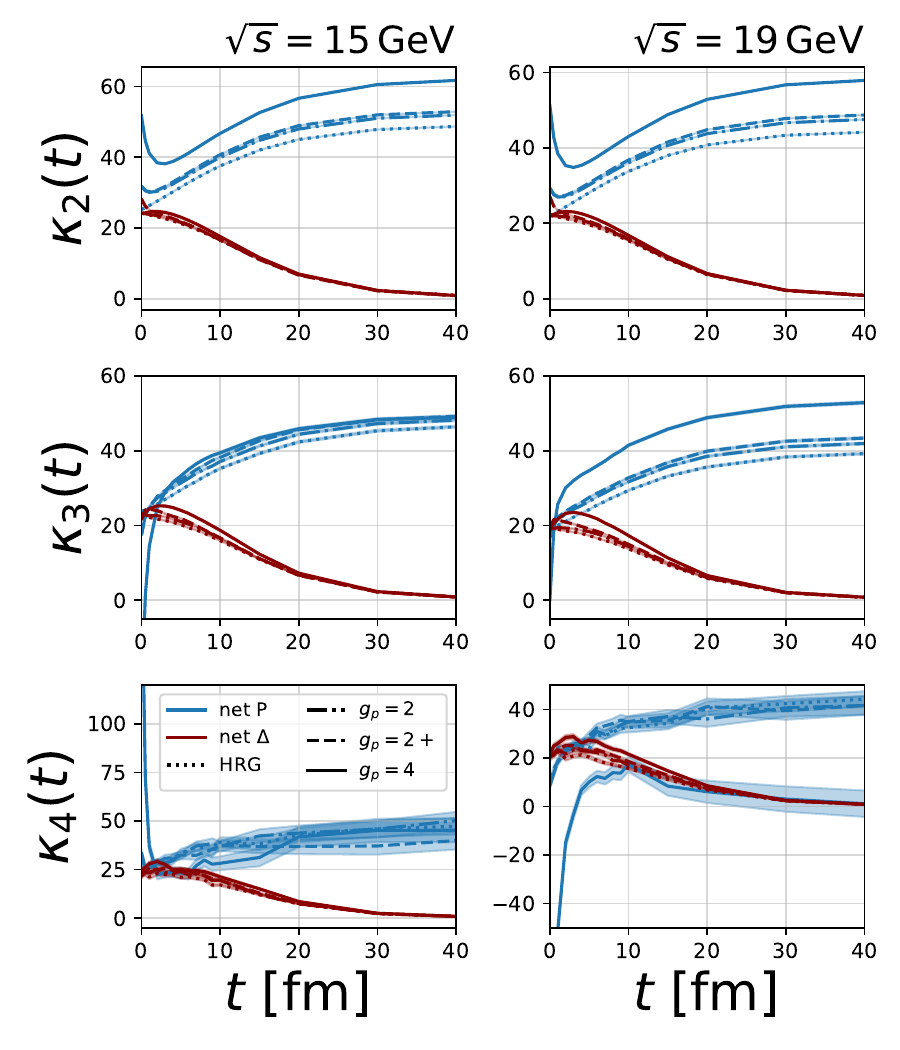}
      \caption{Time evolution of the net proton (blue) and net delta (red) cumulants as a function of time. The second (upper row), third (center) and fourth cumulant (bottom row) are shown for two energies $\sqrt{s} = 15\,\mathrm{GeV}$ (left column) and $\sqrt{s} = 19\,\mathrm{GeV}$ (right column). The results are presented for initializing the system with the cases $g_p = 2$ (dashed-dotted line), $g_p = 4$ (straight line) as well as the case where more hadron species are coupled to the critical field $g_p=2+$ (dashed line) and finally the HRG model (dotted line). A momentum cut of $0.3 < p < 2.0\,\mathrm{GeV}$ is included.}
      \label{Fig:time-evolution}
    \end{figure}
    Fig.~\ref{Fig:time-evolution} shows the evolution of the net proton and net Delta cumulants as a function of time. We have picked those two energies since the strength of the critical point is the strongest in this region. We note that for $\sqrt{s} = 15$ the system is initialized with $\kappa_4^{CP} \gg \kappa_4^{hrg}$ whereas for $\sqrt{s} = 19\,\mathrm{GeV}$ $\kappa_4^{CP} \ll \kappa_4^{hrg}$.
    We also include a cut on the absolute momentum of $0.3 < p < 2$ GeV in order to roughly mimic the experimental situation. Since the delta is the lightest baryonic resonance we are interested in its fluctuations and the interplay between the delta and proton cumulants.
    
    First, the net proton fluctuations initialized with the HRG model show a strong increasing behavior over time for all presented cumulants. From a starting value of $\kappa_2(0)\approx 25$ the the variance nearly doubles to around $\kappa_2(40\,\mathrm{fm})\approx 50$. This is a result of an increasing (anti-)proton number due to the decay of resonances during the evolution.
    
    The $\Delta$-baryon cumulants of all presented orders show the opposite behavior and go to zero over time as they decay into stable particles and therefore the fluctuations vanish.

    We don't observe a large difference between the cases where only nucleons are coupled to the critical mode ($g_p = 2$) or where a larger set of hadrons is coupled ($g_p = 2+$). Even though increased net delta correlations exist in the initial state, they vanish within the first $\approx 5\,\mathrm{fm}$. The difference between the net proton cumulants in the final state of the $g_p = 2$ and the $g_p = 2+$ case is approximately $\approx 2\%$.
    
    Going to the initialization with a stronger coupling to the critical mode $g_p = 4$ a strong modification of $\kappa_n$ within the first $5$ fm towards the HRG case is found within the first $\approx 5\,\mathrm{fm}$. After the first couple of $\mathrm{fm}$ the fluctuations start rising again and the correlations from the critical field are propagated to the final state in both $\sqrt s = 15 \,\mathrm{GeV}$ and $\sqrt s = 19 \,\mathrm{GeV}$.
    
    The evolution of the third cumulant $\kappa_3(t)$ is similar to $\kappa_2(t)$. From the initial value at $t=0$, $\kappa_3$ strongly increases. In the case of negative initial value at $\sqrt{s} = 15\,\mathrm{GeV}$ the signal from the critical point gets completely washed out and the sign of $\kappa_3$ changes. Contrary to $\sqrt s = 15\,\mathrm{GeV}$ we observe that at $\sqrt{s} = 19\,\mathrm{GeV}$ correlations from the critical field survive the hadronic evolution. Similarly to $\sqrt s = 15\,\mathrm{GeV}$ though, $\kappa_3(t)$ starts from $\kappa_3^{critical}(0) \ll \kappa_3^{HRG}(0)$ but ends with correlations $\kappa_3^{critical} \gg \kappa_3^{HRG}$.

    Similarly to the second and third cumulant, the fourth order cumulant is strongly affected within the first couple of fm as well. Here, the strong correlations in the initial state are reduced towards the HRG baseline.
    In the final state for $\sqrt{s} = 15\,\mathrm{GeV}$, the value of $\kappa_4$ evolves towards the HRG evolution and in the final state, no difference within the errors can be observed.
    In the case $\sqrt{s} = 19\,\mathrm{GeV}$ where $\kappa_4(0)\ll 0$ correlations from the critical point survive the hadronic evolution and are present in the final state.
    
    We also observe that e.g. $\kappa_3(t)$ at $\sqrt{s} = 19\,\mathrm{GeV}$ in the case $g_p = 4$ the net delta correlations increases before going down to zero.
    In this case, the $\Delta$-baryons are not coupled to the critical mode meaning that correlations from the net protons are passed between the different particle species in the transport model.

    In the next step, we want to discuss the origin of the modifications of the net proton correlations at the beginning of the expansion. By switching specific interaction channels on and off we can study the dependency of the evolution of the fluctuations on the collision kernel of the transport simulation.
    \begin{figure}[h]
      \centering
      \includegraphics[width=0.45\textwidth]{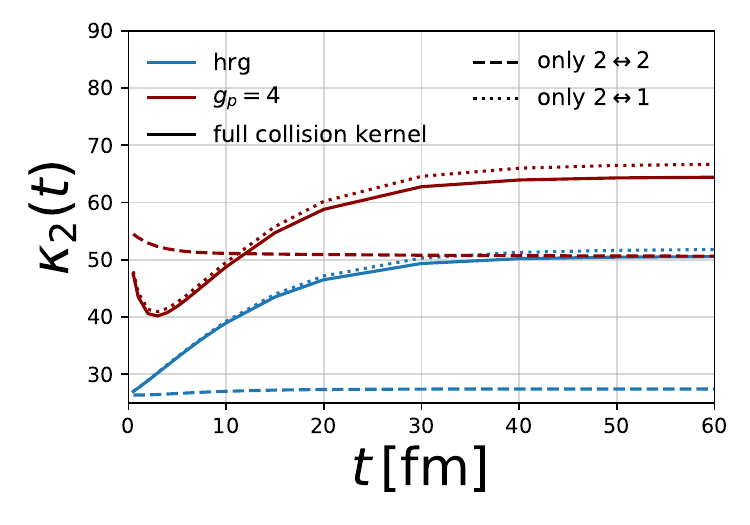}
      \caption{Time evolution of the second net proton cumulants as a function of time. The system is initialized using the HRG  (blue lines) and with a coupling of $g_p = 4$ (red lines). The results of different simulations using the full collision kernel (straight line), employing only $2\leftrightarrow 1$ interactions (dotted line) and (in)elastic scatterings (dashed line) are shown.}
      \label{Fig:time-evolution-collisions}
    \end{figure}
    Fig.~\ref{Fig:time-evolution-collisions} shows the impact of specific interaction channels on the evolution of the net proton number cumulants. 
    We observe that the resonance formation and decays are not only the most occurring interactions during the evolution (see Fig. \ref{Fig:collisions}) but also the ones that have the largest effect on the evolution of net proton cumulants as a function of time.
    
    When switching only to (in)elastic $2\leftrightarrow 2$ interactions $\kappa_2(t)$ only slightly decrease when initialized with a coupling to the critical field. In contrast, when initialized using the HRG model the cumulants slightly increase.
    Now, in the case in which $2\leftrightarrow 2$ interactions are switched off and only $2\leftrightarrow 1$ interactions are considered, we observe a much stronger effect on the cumulants in contrast to performing only $2\leftrightarrow 2$ reactions. Here, we find a similar evolution as in the case of the full collision kernel.
    This shows that the resonance formation/decay processes have the strongest influence on the net proton correlations and the decays are responsible for the increase over time.
    In the case of $g_p = 4$ $\kappa_2(t)$ decreases until a minimum is reached around $t\approx 3\,\mathrm{fm}$. Then, the variance grows again until it saturates.
    The reason for the non-monotonic behavior when initialized with a coupling to the critical field is that within the first stage of the expansion, resonances are created from interactions with a (anti-) proton in the initial state. As a result, the net proton fluctuations decrease in the first 3 fm.
    After the initial resonance formations however the formed unstable particles decay and increase correlations between protons and anti-protons which is responsible for the increase of the variance. This effect is also observed for the HRG initialization.
    It has been checked that the discussed results for the second cumulants also hold for the $\kappa_3$ and $\kappa_4$.
  
  \subsection{Isospin fluctuations}\label{Sec:Isospin}

    We now want to quantify the effect of the dynamical expansion of the hadronic medium on the final state cumulants. It has been shown that the largest sources of fluctuations are resonance formations and decays, which feed into the proton spectra. During the expansion of the medium and especially in the final state there are of course no unstable particles left, but on top of the pure resonance decay process there are many scatterings and resonance regeneration processes that affect the cumulants in a different way.
    It is therefore useful for our purpose to define the following quantity
    \begin{equation}\label{Eq:KappaTilde}
        \tilde\kappa_n = \frac{\kappa_n^{\mathrm{dynamical}}}{\kappa_n^{\mathrm{decays}}} \, .
    \end{equation}
    Here, $\kappa_n^{\mathrm{dynamical}}$ are the final state cumulants after the dynamical expansion of the hadronic medium. On the other hand, we can also directly perform the decays without evolving the medium dynamically and measure the cumulants which are denoted as $\kappa_n^{\mathrm{decays}}$.
    The latter case is similar to calculations performed e.g. in \cite{Garg:2013ata, Nahrgang:2014fza, Bluhm:2016byc}. Pure decay processes alter the cumulants of the net proton cumulant with feed-down processes. As an example, the mean net proton number is modified in the following way
    \begin{equation}
        \kappa_1 = \langle N_p \rangle - \langle N_{\bar p} \rangle + \sum_R  \langle N_R \rangle (\langle n_p\rangle_R - \langle n_{\bar p}\rangle_R ) \, .
    \end{equation}
    Here, $\langle n_p\rangle_R = \sum_r b_r^R n_{i,r}^R$ is the average number of protons originating from all decay channels with branching ratios $b_r^R$. Within the transport code the above equation is performed on a Monte Carlo basis where additional fluctuations arise due to the mass-dependent decay width $\Gamma(m)$ of each resonance and their sampled masses from the thermal spectral function $\mathcal{A}(m)$.
    
    \begin{figure}[h]
      \centering
      \includegraphics[width=0.35\textwidth]{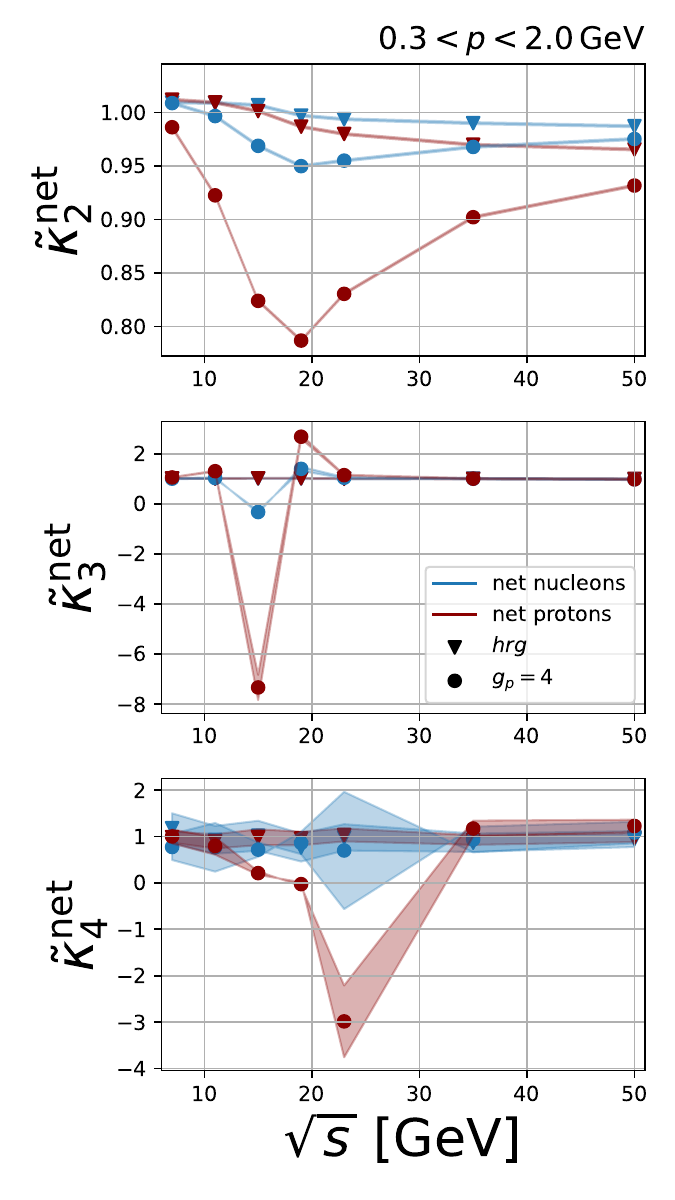}
      \caption{$\tilde\kappa_n$ as a function of $\sqrt{s}$ including a cut in momentum space for net proton (red) and net nucleon (blue) fluctuations. The results are show for the hrg initialization (triangles) and with a coupling of $g_p = 4$ to the critical field (circles).}
      \label{Fig:kappa-tilde}
    \end{figure}
    Fig.~\ref{Fig:kappa-tilde} shows the results of Eq. \ref{Eq:KappaTilde} as a function of the $\sqrt{s}$.
    In the case of initialization with the HRG model, we observe a difference from unity in the second net proton cumulant $\tilde\kappa_2$ where a suppression of the dynamically evolved cumulant of $\approx 2\%$ is observed. A large portion of this suppression originates from isospin randomization processes since $\tilde\kappa$ of nucleons is not strongly affected. For example, the process $p\pi^0 \leftrightarrow \Delta^+ \leftrightarrow n\pi^+$ modifies the proton number whereas the nucleon number is not affected.
    Within the errors, no difference from unity can be observed in the ratios $\tilde\kappa_3$ and $\tilde\kappa_4$.
    
    If the system is initialized with a coupling of nucleons to the critical mode larger modifications of $\tilde\kappa$ are observed.
    Similar to the HRG initialization on the level of the second cumulants a suppression of $\tilde\kappa_2$ of the final state cumulants after the dynamical evolution with respect to performing only the decays is seen. However, in this case the suppression is on the order $20\%$ at $\sqrt{s} = 19\,\mathrm{GeV}$. As already described in the HRG case the suppression originates from isospin randomization processes as the net nucleon fluctuations are less affected.
    
    For the third and fourth cumulant, a strong modification in $\tilde\kappa_n$ is observed. The expansion of the medium at $\sqrt{s} = 15\,\mathrm{GeV}$ evolves the third net proton cumulant from a value $\kappa_3\ll 0$ towards $\kappa_3\gg 0$ (see e.g. Fig. \ref{Fig:time-evolution}) whereas the evolution using only decays preserve the negative skewness from the initial state. As a result, the ratio $\tilde\kappa_3$ changes its sign around the region where the signature of the critical point is the strongest.
    
    We find that the dynamical evolution modifies the net proton $\kappa_4$ in a similar way compared to the third cumulant. In the scaling region of the critical point, the strong correlations get washed out by the hadronic interactions whereas performing only decays preserve these correlations. We additionally see that at $\sqrt{s} = 15\,\mathrm{GeV}$ and $\sqrt{s} = 19\,\mathrm{GeV}$ where the initial values are either $\kappa_4^{\mathrm{initial}}\gg 0$ or $\kappa_4^{\mathrm{initial}}\ll 0$ the cumulants in the final state are differently strong affected. The dynamical evolution washes out a strong positive initial $\kappa_4$ towards the HRG baseline whereas performing the resonance decays preserves these initial correlations. The sign change in $\tilde\kappa_4$ originates from the fact that initial $\kappa_4^{\mathrm{initial}}\ll 0$ evolve towards positive $\kappa_4$ above $\sqrt{s} = 19\,\mathrm{GeV}$. Again, this change can be attributed to the isospin randomization processes as described above.

  \subsection{Final state observables}\label{Sec:FinalState}
    We now want to present ratios of the final state cumulants after the full dynamical evolution of the hadronic medium. We show the results with the same cut in momentum space as before, again in order to mimic experimental constraints. Since this work is not in the stage of making any comparisons to experimental measurements we don't include the data points here.
    \begin{figure}[h]
      \centering
      \includegraphics[width=0.5\textwidth]{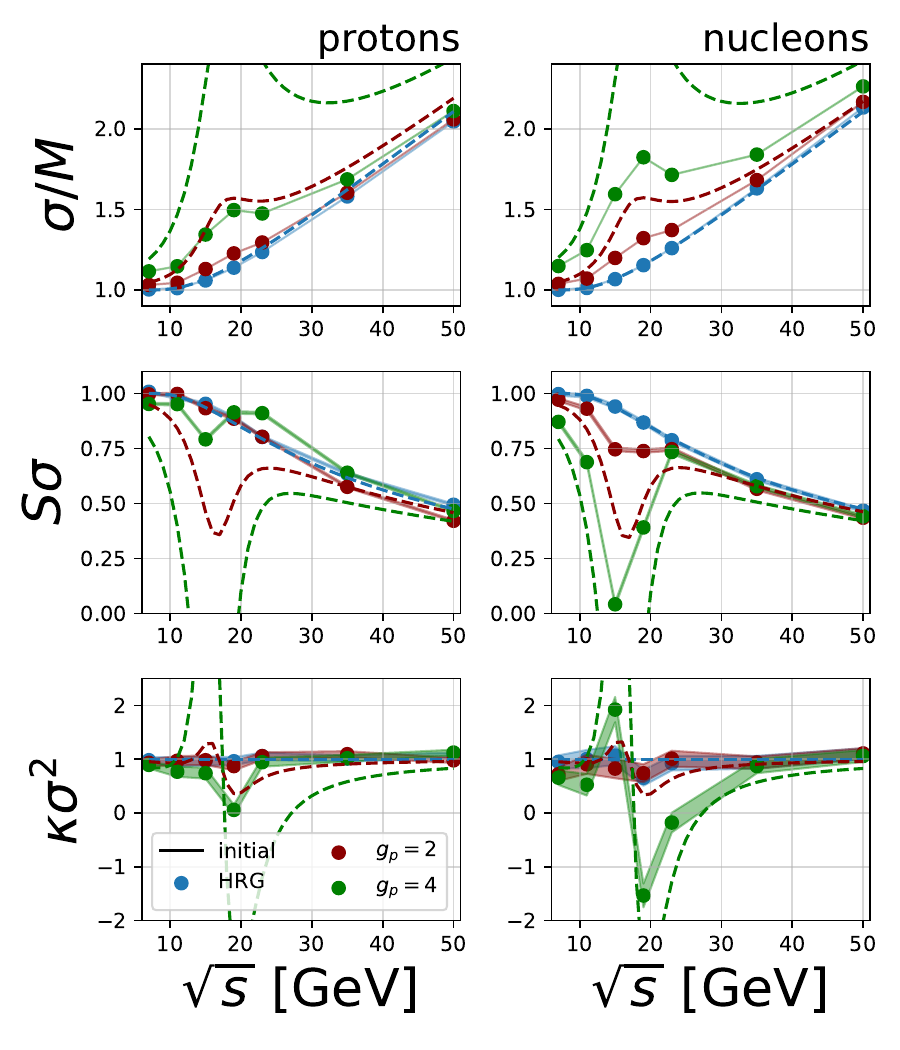}
      \caption{Ratios of final state proton (upper row) and nucleon (lower row) cumulants as a function of $\sqrt{s}$. The scaled variance (left column) and skewness (center) kurtosis (right column) are shown. The results from SMASH are shown as points for the hrg initialization (blue) and coupling nucleons to the critical field with $g_p = 2$ (purple) and $g_p = 4$ (yellow) whereas the analytic results from the initial state are shown as dashed lines.}
      \label{Fig:final-state}
    \end{figure}
    
    Fig.~\ref{Fig:final-state} shows the scaled variance skewness and kurtosis of the final state after the dynamical evolution.
    Starting with the HRG initialization we can see that both the net proton and net nucleon fluctuations are on top of the analytic initial state line. This means that the previously discussed effects of resonance regeneration processes and isospin randomization equally affect $\kappa_{2,3,4}$. As a result, the ratios are in line with the HRG expectation.
    
    In the case of $g_p = 2$, we find that the net proton cumulants of the final state only contain very little correlation from the critical point. $\sigma / M$ shows a slight enhancement near the scaling region of the CP. Higher order cumulants however show no signs of the initial correlations.
    Due to fewer isospin fluctuations, the nucleon fluctuation contains more correlations in the final state in $\sigma / M$ and $S\sigma$.
    
    If the system is initialized with a stronger coupling of $g_p = 4$ we observe that for all presented ratios of net proton cumulants, correlations from the critical point point survive the evolution of the hadronic medium, even though these correlations are much weaker than the ones from the initial state.

    The ratio $\kappa\sigma^2$ shows a non-monotonic behavior similar to the initial state in comparison to the baseline curve.
    At $\sqrt{s} = 15$ GeV where $\kappa\sigma^2_{\mathrm{initial}}\gg 1$ the final state value is $\approx 1$ whereas at $\sqrt{s} = 19$ GeV where $\kappa\sigma^2_{\mathrm{initial}}\ll 1$ the final state value is $\approx 0$.
    Similarly to $g_p = 2$, the net nucleon fluctuations show much larger correlations in the final state in comparison to the net proton cumulants. Even though the final state values are smaller in comparison to the initial state, the shape is very similar to what was put into. As a result and similar to our previous conclusions we find that isospin randomization processes greatly impact the evolution of the cumulants.
  
  \subsection{Rapidity dependence}\label{Sec:RapidityDependence}
    In this section the dependency of the fluctuations as a function of the rapidity window $\Delta y$ with $y = \frac{1}{2}\mathrm{log}((E+p_z) / (E - p_z))$ is investigated. This is important for comparison with experimental measurements where the cumulants are measured in momentum space and not the full phase space can be observed \cite{Ling:2015yau, Brewer:2018abr}.
    We note that the dynamics of the system considered here are not the same as in heavy-ion collisions since no distinct direction exists in the expanding sphere. However, for a comparison and in order to define the cuts in momentum space, we employ the definition of rapidity and $p_T$ in our setup. The starting point is the integrated net proton density $n^{\mathrm{net}}(y, p_T)$ over a given rapidity and $p_T$ interval
    \begin{equation}
      N^{\mathrm{net}}(\Delta y) = \int_{-\Delta y/2}^{\Delta y/2}dy\int_{0.3}^{2}dp_T\, n^{\mathrm{net}}(y, p_T) \, .
    \end{equation}
    From here, the cumulants $\langle (\delta N^{net}(\Delta y))^n\rangle$ are calculated in each rapidity interval and are shown as a function of $\Delta y$.

    \begin{figure}[h]
      \centering
      \includegraphics[width=0.5\textwidth]{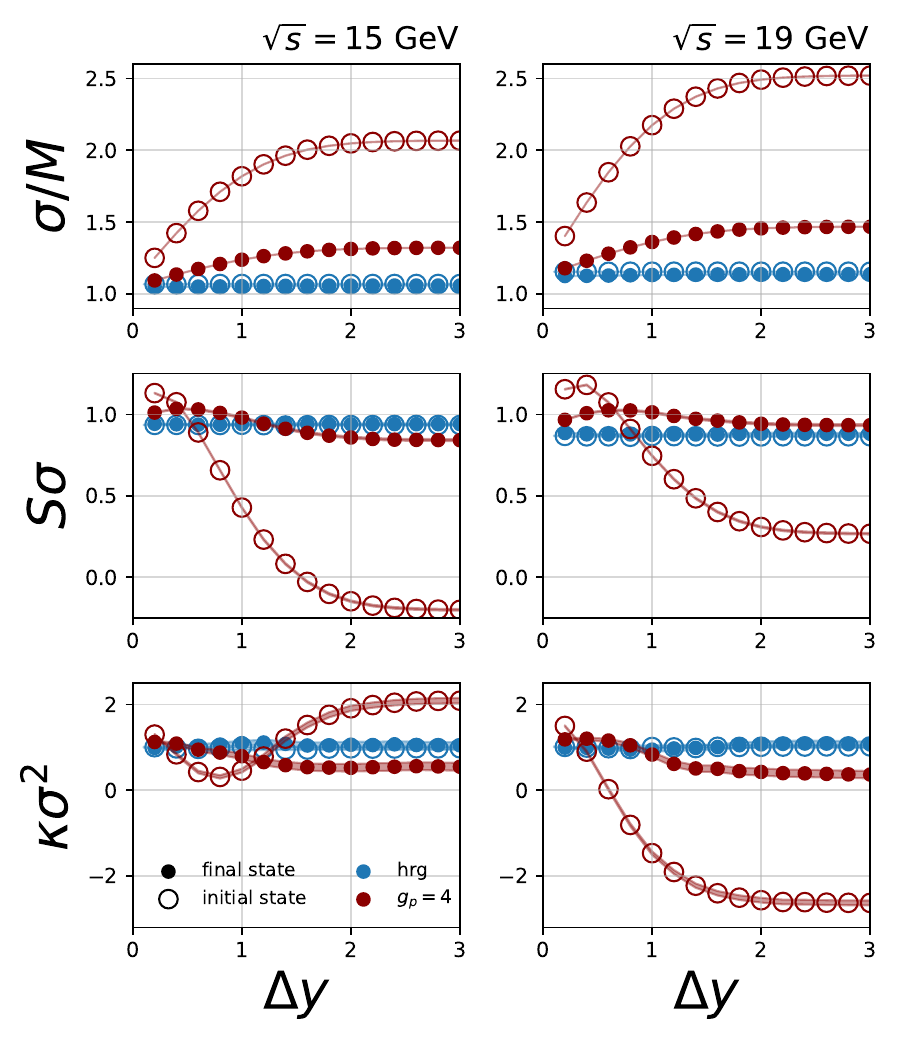}
      \caption{Net proton scaled variance (top), skewness (center) and kurtosis (bottom) as a function of the rapidity window $\Delta y$ for $\sqrt{s} = 15\,\mathrm{GeV}$ (left) and for $\sqrt{s} = 19\,\mathrm{GeV}$ (right). The results of the initial (open circles) and final state (closed circles) are shown for the HRG (blue) and $g_p = 4$ (red) case.}
      \label{Fig:DeltyY}
    \end{figure}
    Fig.~\ref{Fig:DeltyY} shows the initial and final state ratios of cumulants as a function of the rapidity interval $\Delta y$. For the HRG results no rapidity dependence is observed for both the initial and final state cumulants. This is expected as the system is initialized according to the grand canonical ensemble and therefore no correlations appear along the momentum directions. It is expected that the effects of exact charge conservation have a strong influence on the rapidity dependence \cite{Bzdak:2012an, Hammelmann:2022yso}. 
    When initialized with a coupling to the critical field a difference between initial and final state is observed. 
    The scaled variance shows an increasing behavior with larger rapidity windows until it reaches a plateau. The difference between the initial and final state fluctuations is only in magnitude, whereas the shape is similar.
    The skewness starts at $S\sigma > 1$ at small with a positive slope at small rapidity windows before decreasing. After the dynamical evolution, the strong correlations of the critical point vanish however the shape of the rapidity dependence is similar at the initial and final state. However, the maximum appears to grow towards a larger rapidity window. Similar to what has been shown in Fig.~\ref{Fig:final-state}, the net proton skewness in the full rapidity window is above and below 1 for $\sqrt{s} = 15\,\mathrm{GeV}$ and $\sqrt{s} = 19\,\mathrm{GeV}$, even though $S\sigma\ll 1$ in the initial state.
    The kurtosis shows a similar behavior in the final state of the evolution as the skewness. The kurtosis is above unity for small rapidity windows $\Delta y < 0.5$ before decreasing to $\kappa\sigma^2 < 1$ at large rapidity windows. At $\sqrt{s} = 15\,\mathrm{GeV}$ and compared to the initial state correlations the behavior of the final state kurtosis is opposite to the initial state. Similar to the skewness, the kurtosis, when initialized with $\kappa\sigma^2\gg 1$, evolves towards $\kappa\sigma^2 < 1$.
    From Fig.\ref{Fig:DeltyY} one can directly observe the importance of the applied momentum cuts. Within this model, the strong correlations from the critical point appear at large acceptances whereas when one only looks at small rapidity windows the fluctuations yield different results.
    
\section{Conclusion and Outlook}\label{Sec:Conclusion}
    We have studied the influence of hadronic interactions on the evolution of critical fluctuations in a transport model. We have shown that the maximum entropy distribution successfully reproduces the first four cumulants from the HRG coupled to the 3d Ising model. We then established the initial state in coordinate and momentum space and studied the expanding medium's thermodynamic evolution and collision types.
    In the first part, we investigated the time dependence of the cumulants and found that resonance formation and decay processes have the strongest influence on the fluctuations. We then quantified the impact of isospin randomization processes and the final state net proton and net nucleon cumulants as a function of $\sqrt{s}$. We have shown that in the case of a coupling $g_p = 4$ correlations from the critical point survive the hadronic evolution. We have also studied the rapidity dependence of the final state fluctuations.

    For future work, it would be interesting to study more realistic heavy-ion collision scenarios and incorporate the described procedure in a Cooper-Frye sampler for hydrodynamic calculations.
    We also note that the transport model used to evolve the hadronic medium does not propagate the n-particle correlations but only the single-particle distribution function. It would therefore be important to incorporate the effects of a critical point on the level of mean-field potentials in the transport model similar to \cite{Sorensen:2020ygf}.
    
\appendix
\section{Thermal fits}\label{App:1}
  The thermal equilibrium values of the temperature and chemical potentials are obtained by performing a thermal model fit. Here the multiplicities of the following stable particles $N, \pi, K, \Sigma, \Lambda$ plus their respective anti-particles are used. Together with Eq.~\ref{Eq:hrg_density} one than minimizes the following function to obtain the thermodynamic quantities plus the volume of the system
  \begin{equation}
    \begin{aligned}
    \chi(T, V, \mu_{B,Q,S}) = & \frac{1}{N_{\mathrm{species}}} \sum_i^{N_{\mathrm{species}}} \\ 
                                      & (N_i^{\mathrm{SMASH}} - V n_i(T, \mu_{B,Q,S}))^2 \, .
    \end{aligned}
  \end{equation}

  \begin{figure}[h]
    \centering
    \includegraphics[width=0.5\textwidth]{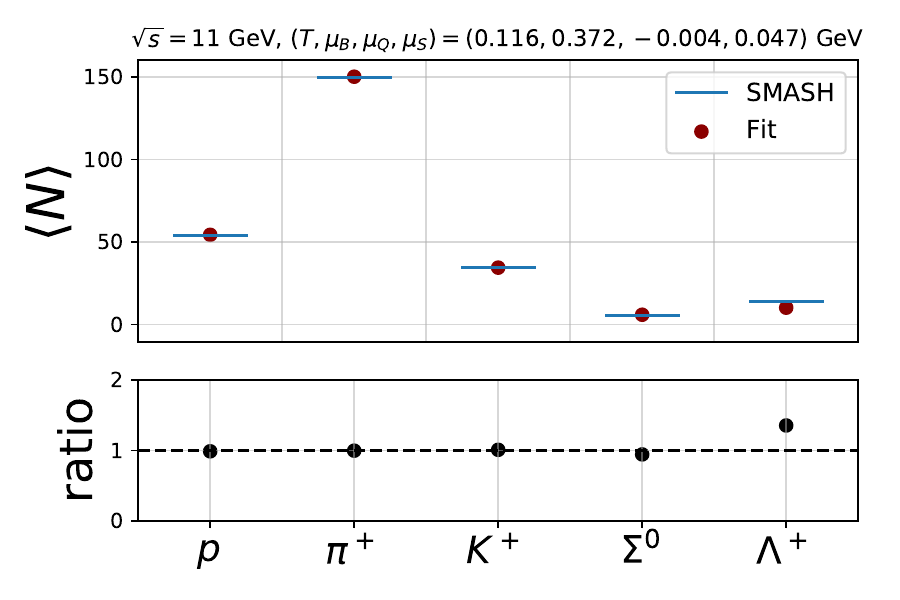}
    \caption{Yields from SMASH (blue line) and thermal model fit (red points) for $\sqrt{s} = 11\,\mathrm{GeV}$ of the most abundant stable particles.}
    \label{Fig:ThermalFit}
  \end{figure}
  One such a fit is shown in Fig.~\ref{Fig:ThermalFit} for the $\sqrt{s} = 11\,\mathrm{GeV}$. One can see that for the presented particles the fit well reproduces the number of particles with the exception of the anti-protons. It is not clear at this point where this difference comes from.

\section{Particles coupled to the critical mode}\label{App:2}
  The list of particles that are coupled to the critical field is shown in Tab.~\ref{App:TableDoF}.
  \begin{table}[hbt!]
    \begin{tabular}{| c | c | c | c |}
      \hline
      Particle & Mass $\mathrm{[GeV / c^2]}$ & Degeneracy \\\hline\hline
      $\pi$ & $0.138$ & $3$ \\\hline
      $\rho$ & $0.776$ & $6$ \\\hline
      $K$ & $0.494$ & $4$ \\\hline
      $K^\star(892)$ & $0.892$ & $8$ \\\hline
      $N$ & $0.938$ & $8$ \\\hline
      $\Delta$ & $1.232$ & $32$ \\\hline
      $\Lambda$ & $1.116$ & $2$ \\\hline
      $\Sigma$ & $1.189$ & $12$ \\\hline
    \end{tabular}
    \caption{Hadronic degrees of freedom used in the presented calculations. The degeneracy is the product of spin, charged and anti-particle states.}
    \label{App:TableDoF}
  \end{table}

\begin{acknowledgments}
  This work was supported by the DFG SinoGerman project - Project number 410922684. Computational resources have been provided by the Center for Scientific Computing (CSC) at the Goethe-University of Frankfurt. H.E. acknowledges the support by the State of Hesse within the Research Cluster ELEMENTS (Project ID 500/10.006). This work was in parts supported by the European Union's Horizon 2020 research and innovation program under grant agreement number 824093 (STRONG-2020).
\end{acknowledgments}

\bibliography{bibliography.bib}

\end{document}